\documentclass[aps,prl,reprint,preprintnumbers]{revtex4-1}

\usepackage{amsmath}
\usepackage{amsfonts}
\usepackage{amssymb}
\usepackage{graphicx}
\usepackage{xcolor}

\newcommand\half{\hbox{$\frac12$}}
\newcommand\third{\hbox{$\frac13$}}
\newcommand\fourth{\hbox{$\frac14$}}

\begin{document}
\preprint{MIT-CTP/4926}

\title{Regularizations of Time Crystal Dynamics}


\author{Alfred D. Shapere$^1$, Frank Wilczek$^{2,3,4,5,6}$}
\affiliation{$^1$Department of Physics and Astronomy, University of Kentucky, Lexington, Kentucky 40506 USA\\
$^2$Center for Theoretical Physics, Massachusetts Institute of Technology, Cambridge, Massachusetts 02139 USA\\
$^3$T. D. Lee Institute, Shanghai, China\\
$^4$Wilczek Quantum Center, Department of Physics and Astronomy, Shanghai Jiao Tong University, Shanghai 200240, China\\
$^5$Department of Physics, Stockholm University, Stockholm SE-106 91 Sweden\\
$^6$Department of Physics and Origins Project, Arizona State University, Tempe AZ 25287 USA}

\date{\today}

\begin{abstract}
We demonstrate that non-convex Lagrangians, as contemplated in the theory of time crystals, can arise in the effective description of conventional, physically realizable systems.  Such embeddings resolve dynamical singularities which arise in the reduced description.  Microstructure featuring intervals of fixed velocity interrupted by quick resets -- ``Sisyphus dynamics '' --   is a generic consequence.  In quantum mechanics this microstructure can be blurred, leaving entirely regular behavior. 
\end{abstract}


\maketitle


\bigskip

The concept of ``time crystal'' \cite{classicalTXTal,Wilczek:2012jt,Sacha:2017fqe} has attracted great interest recently, both theoretical \cite{Sacha2015,Smolyaninov2015eya,Khemani2015,keyserlingk2016a,keyserlingk2016b,Khemani2016yjj,else2016,Else:2017ghz,Yao2017} and experimental \cite{Li2012,Zhang2016,Choi2016}.   Most of the recent activity concerns many-body physics, and the possibility to break time translation symmetry in ways that retain as much as possible of the structure which physicists have come to associate with spontaneous symmetry breaking in other contexts, such as sharp phase transitions, order parameters, and generalized rigidity (e.g. \cite{anderson}).   Here we will explore a different but complementary aspect: effective dynamics.

To be concrete, let us consider adding a potential to the minimal time crystal Lagrangian for a single degree of freedom \cite{classicalTXTal}:
\begin{equation}\label{tXtalL}
L ~=~ \frac{1}{12} {\dot y}^4 - \frac{1}{2} {\dot y}^2 - V(y).
\end{equation}
Aside from its connection to time crystals, we can motivate Eqn.\,(\ref{tXtalL}) in the spirit of Landau's philosophy of effective field theory, wherein one considers the coefficients of plausible interaction terms -- in practice, low order polynomials -- as parameters, which can vary with external conditions such as temperature.  Often, interesting behaviors -- changes in phase, or in pattern - arise when the coefficient of some term changes sign.    Conventionally this level of generality is applied to potential energy terms, but in principle one should bring in kinetic energy terms.  This leads us to Eqn.\,(\ref{tXtalL}), as the simplest nontrivial example.

This Lagrangian leads to the energy -- that is, the constant of the motion which is connected by Noether's theorem to time translation symmetry --
\begin{equation}\label{tXtalE}
E = \frac{1}{4} ({\dot y}^2 -1 )^2 + V(y) - \frac{1}{4}. 
\end{equation}
If $V(y)$ has an isolated minimum, then minimizing this energy leads, on the face of it, to a mathematical contradiction.  Indeed, minimizing the potential energy leads us to a fixed value of $y$, while minimizing the kinetic energy leads us to a non-zero velocity.   No regular function can have both a fixed value and non-vanishing derivatives.   Note that similar mathematical problems arise in a purely spatial context, if we try to minimize following energy function for a one-dimensional system: 
\begin{equation}
E_{\rm spatial} ~=~ \int dx \, \Bigl[\Bigl(\bigl(\frac{\partial \phi}{\partial x}\bigr)^2 - 1\Bigr)^2 + U(\phi) \Bigr]\, . 
\end{equation}
Energy integrals of this type are the subject of a substantial mathematical literature \cite{singularVariational}, and also arise in models of materials \cite{muller1999,kohn2006}.  Our treatment of the time crystal problem suggests new possibilities in those areas, as we shall discuss further below.

We can gain a more general perspective by considering not only the (problematic) ground state but solutions of the equations of motion more generally.  
In the equation of motion
\begin{equation}\label{tXtalEofM}
( {\dot y}^2 -1 )\ddot y = - V^\prime (y).
\end{equation}
we see that the effective mass, $\dot y^2-1$, can vanish and change sign.  Negative effective mass is unusual, though perhaps not problematic in itself,  at the level of differential equations.  But vanishing effective mass, in the framework of Newtonian mechanics, signals that the evolution equation becomes either trivial or ill-defined.   For that reason, one might be inclined to think that the behavior implied by Eqn.\,(\ref{tXtalL}) is inherently pathological, and physically unrealizable. 

Here however we will demonstrate that, to the contrary, Eqn.\,(\ref {tXtalL}) arises as the effective description of a realistic physical system in an appropriate limit.   The realization implies a specific regularization of the singular behavior.  As the limit is approached, Eqn.\,(\ref {tXtalL}) governs the behavior of the system accurately except during brief intervals.   

More specifically, let us suppose that $V(y) = \frac{1}{2} y^2$ and that we choose the energy near its absolute minimum $-\frac{1}{4}$.  Then, as long as Eqn.\,(\ref {tXtalE}) applies, the system must have velocity near $\dot y = 1$ (or $-1$), yet stay close to $y = 0$.   It can do this most of the time, if during very brief intervals the regulator comes into play, and allows a quick transit between small positive and small negative values of $y$.   Alluding to the famous myth, we call this ``Sisyphus dynamics''.  It is the behavior we will find to occur.   

Thus, we discover that close to its energy minimum, solutions of the minimal time crystal Lagrangian, appropriately regulated, feature characteristic low-amplitude, high-frequency oscillations.   The system behaves, in other words, as a tunable, non-dissipative relaxation oscillator, exhibiting temporal microstructure.  Here we analyze concrete problems involving  a charged particle in special magnetic and electric fields, but since the underlying mathematical mechanism is simple and general, we anticipate that Sisyphus dynamics will emerge, through the same mathematical mechanism, in other 
contexts.   


\bigskip

{\it Model: Planar charge in external fields.\/} \ Consider the Lagrangian
\begin{equation}\label{regulatedL}
L ~=~ \frac{\mu}{2} {\dot x}^2 + f(x) \dot y - g(x) - V(y).
\end{equation}
This corresponds to a planar charged particle subjected to the magnetic field $B_z = f^\prime(x)$ and the electric potential $g(x) + V(y)$. (For simplicity, we assume here an asymmetric mass parameter, which vanishes in the $y$ direction.) 
We have the equations of motion
\begin{eqnarray}
\mu \ddot x &=& f^\prime (x) \dot y - g^\prime(x) \label{xEquation} \\
\dot x f^\prime(x) &=& - V^\prime(y).  \label{yEquation}
\end{eqnarray}

Now let us consider the idealization $\mu \rightarrow 0$, which can be appropriate for strong magnetic fields.   Then we have, from Eqn.\,(\ref{xEquation}), formally
\begin{equation}\label{yDotEquation}
\dot y = g^\prime / f^\prime. 
\end{equation}
Choosing $f(x) = \frac{1}{3} x^3 - x$, $g(x) = \frac{1}{4} x^4 - \frac{1}{2} x^2$, this becomes simply $\dot y = x$.  Replacing $x$ by $\dot y$ in the remaining equation of motion Eqn.\,(\ref{yEquation}) then reproduces Eqn.\,(\ref{tXtalEofM}). 

Alternatively, we can use Eqn.\,(\ref{xEquation}) to eliminate $x$ from
 Eqn.\,(\ref{regulatedL}) (with $\mu = 0$) to arrive at Eqn.\,(\ref{tXtalL}) directly.  (This demonstrates, {\it inter alia}, that our neglect of the mass parameter in the $y$ direction is inessential in the time crystal regime, since Eqn.\,(\ref{tXtalL}) already includes a term of the form it generates.  Including such a term explicitly shifts the critical velocity from $1$ to $\sqrt{1-\mu}$.   More notably, by making $y$ dynamical we also enlarge the phase space.  With that enlarged space, the time crystal effective theory of (1) governs a robust but limited range of choices of initial conditions for $x$ and $y$.)
 
The sensitive point in this derivation is that in ``deducing'' Eqn.\,(\ref{yDotEquation}) we will, when $f^\prime = 0$, have divided by zero.  Physically, this occurs at points where the magnetic field vanishes.  At such points we cannot neglect the mass $\mu$, even if it parametrically small.   

Conversely, we see that including a small positive $\mu$ acts as a regulator for the dynamical system defined by Eqn.\,(\ref{tXtalL}).  In this way, we have realized the minimal classical time crystal, including a potential, with a well-defined regulator, as a reduced description (effective theory) of a reasonably simple, physically realistic dynamical system.  When we pass to quantum mechanics, below, this regulator can be removed.   It is an interesting question, whether there are alternative, significantly different regulators of comparable simplicity. 

\bigskip

{\it Sisyphus Dynamics: Microstructure and Ratcheting.\/} \  Our system 
\begin{eqnarray}\label{dynamicalSystem}
\mu \ddot x &=& (x^2 -1)  (\dot y - x)  \label{xEquation2} \\
\dot x (x^2-1) &=& - y \label{yEquation2} \label{giraffe}
\end{eqnarray}
(with $V(y) = \half y^2$) is readily amenable to numerical study, which proves very revealing.  
Note that we can put this system into a more conventional form by using the time derivative of the second equation to eliminate $\dot y$ from the first.

Before we display that characteristic behavior graphically, some interpretive comments are in order: 

\begin{itemize}

\item The initial value problem is well-posed with the specification of $x$ and $\dot x$ at some initial time.  In this formulation, $y$ is a particularly interesting dependent variable, because we expect that it should reflect the time crystal dynamics directly.  

\item The energy function is 
\begin{equation}
E ~=~  \frac{\mu}{2} {\dot x}^2 + \frac{1}{4} x^4 - \frac{1}{2} x^2 + \frac{1}{2} y^2.
\end{equation}
It is minimized by $x = \pm 1$ and $y=0$, independent of time.  At the minima, the energy is $E = - \frac{1}{4}$.   

\item The characteristic ``time crystal'' temporal microstructure arises when the energy approaches, but is not equal to, that minimum value.   Figure 1, the result of a numerical calculation using {\it Mathematica}, displays that behavior graphically.   Note that the contribution $\frac{1}{2} \mu {\dot x}^2$ of the regulator to the energy is very small.  

\item As one considers solutions whose energy approaches the minimum, the frequency of the oscillations in $y$ increases while their amplitude decreases.   The approach to the limit is, qualitatively, highly nonuniform.  Quantitatively, though it is continuous in the norm ${\rm Max} \, | y |$, it is not continuous (for instance) in the norm ${\rm Max} \,  |y| + {\rm Max} \, |{\dot y}|$, nor in simple Sobolev norms.

\end{itemize}

\vspace{0.1in}
\begin{figure}[ht]
\includegraphics[scale=.8]{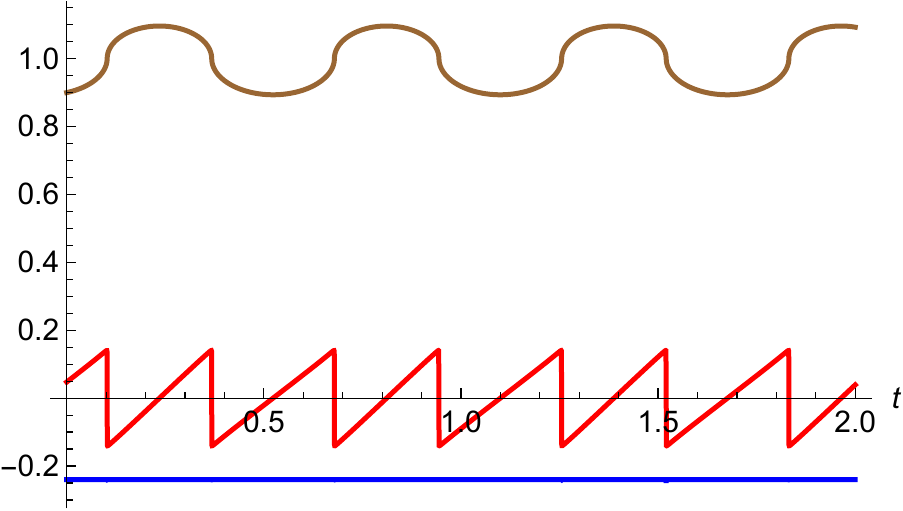}
\caption{Numerical solution of Eqns.\,(\ref{dynamicalSystem},\ref{giraffe}) with $\mu = 10^{-5}, \, x(0) = 0.9, \, {\dot x} (0)  = 0.25\,$.  The upper (gold) curve represents $x(t)$; the middle (red) curve represents $y(t)$; and the lower (blue) curve represents $E$.  The behavior of $y(t)$ exhibits the characteristic temporal microstructure discussed in the text, with near-constant velocity often, but briefly, interrupted by small, sudden jumps.}
\end{figure}

More generally: Within the effective theory, a significant energy barrier separates the positive velocity from the negative velocity region in velocity space.  Thus, when the energy is too small to bridge the gap, the velocity will maintain a constant sign. But that leads, as before, to trouble with the potential energy.  So we might expect, in this more general situation, that the velocity is almost always positive (or almost always negative), interrupted by brief intervals when the effective theory breaks down, and the position gets reset.  That behavior is indeed evident in the numerical simulations, as exemplified in Figure 2.   Prominent in Figure 2, but also subtly present in Figure 1, is a diphasic structure in $y(t)$: the positive velocity evolves with two distinct patterns, depending upon whether $x$ is greater than or less than unity.

\vspace{0.1in}
\begin{figure}[ht]
\includegraphics[scale=.8]{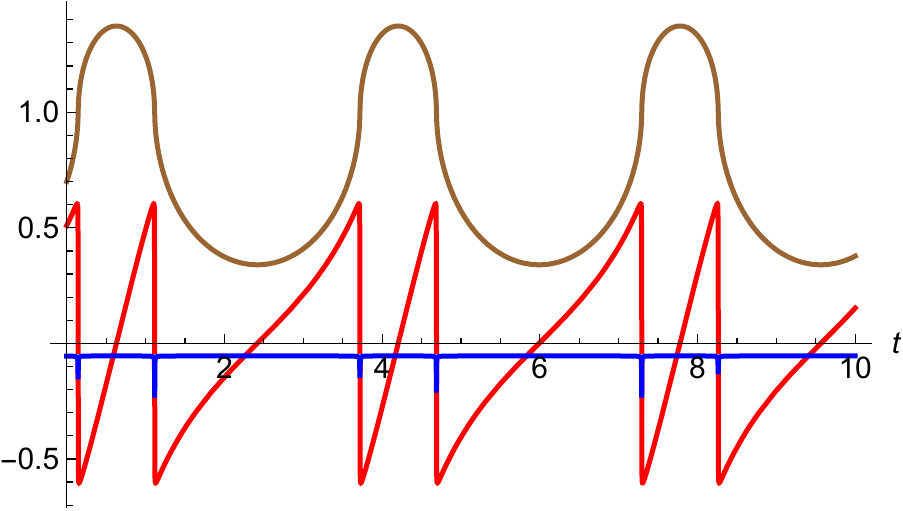}
\caption{Numerical solution of Eqns.\,(\ref{dynamicalSystem},\ref{giraffe}) with $\mu = 10^{-3}, \, x(0) = 0.7, \, {\dot x} (0)  = 1.0\,$.    The color scheme is the same as in Figure 1, except that the blue curve now displays the total energy minus the regulator contribution $\frac{\mu}{2} {\dot x}^2$.  (In Figure 1, this was indistinguishable from the total energy.)  The behavior of $y(t)$ exhibits the characteristic ratcheting discussed in the text, with intervals of positive velocity interrupted by sudden jumps. Note the spikes in the blue curve during the jumps in $y(t)$; this behavior, reflecting regulator contributions to the energy, is more visible in Figure 3.}
\end{figure}

It is also instructive to consider Figure 3, which displays a numerical solution near the energy minimum with a much larger value of the regulator.  We see similar qualitative features -- incipient Sisyphus dynamics -- but with less abrupt switching.

\bigskip

\vspace{0.1in}
\begin{figure}[ht]
\includegraphics[scale=.8]{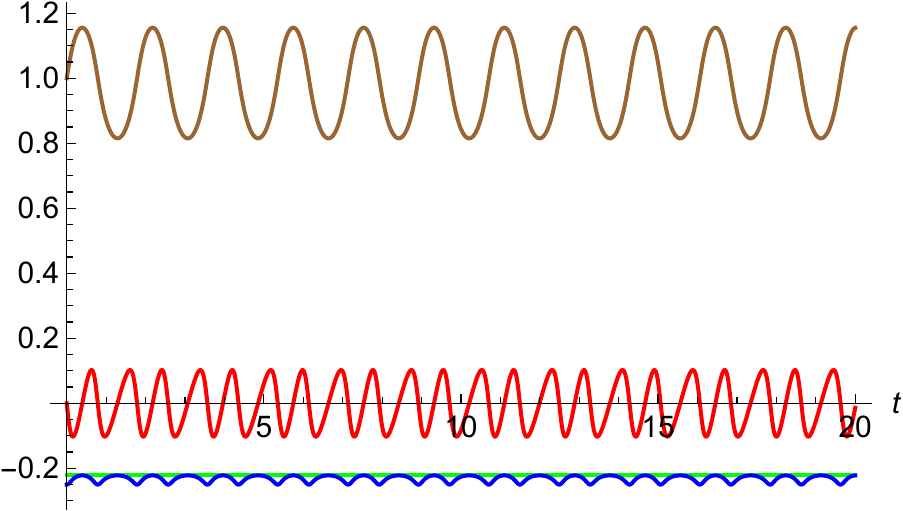}
\caption{Numerical solution of Eqns.\,(\ref{dynamicalSystem},\ref{giraffe}) with $\mu = .1,  x(0) = 1.0,  {\dot x} (0)  = .95$.  The color scheme is the same as in Figure 2, with a horizontal light green line now indicating the total energy.   The behavior of $y(t)$ exhibits the characteristic ratcheting discussed in the text, with intervals of positive velocity interrupted by jumps, but the jumps are now less abrupt.}
\end{figure}

{\it Quantization\/}: \ Suppression of small-amplitude, high-frequency oscillations, in Planck's theory of black-body radiation, was the first mission of quantum theory.  Thus, it is appropriate to explore the quantum version of our model, to see how quantization reflects and modifies the (classical) temporal microstructure. 


Since the classical singularity occurs near the energy minimum, the most salient issue is the spectrum of bound states localized near the potential minimum.  A first, heuristic step is to consider the semiclassical Bohr-Sommerfeld condition.  It proves convenient to write this in its phase-space form, according to which the area $A_n$ associated with energies less than the $n^{\rm th}$ eigen-energy $E_n$ is approximately $2\pi \hbar n$, for $n$ not too small.   We have evaluated this condition in both the original time crystal ($y$) picture and the regulated ($x$) picture for small $\mu$, assuming energy close to the classical minimum $-\frac{1}{4}$, with the concordant result
\begin{equation}\label{semiClassicalEigenvalues}
E_n + \frac{1}{4} ~=~ \left(\frac{3\pi\hbar n}{4\sqrt 2}\right)^{\frac{2}{3}}.
\end{equation}
Similar reasoning can be used to show that large, positive energy levels scale with a different power of $n$:  $E_n\sim n^{4/5}+{\cal O}(1)$ as $n\to \infty$. 

For rigorous quantization \cite{Branched2012}, we must pass to the regulated theory, and to a Hamiltonian formulation.  For that purpose it is convenient to add a total derivative to Eqn.\,(\ref{regulatedL}), so that $f(x) \dot y \rightarrow - y f^\prime (x) \dot x$, and express everything in terms of $x$.   Thus we find
\begin{equation}\label{regularizedHamiltonian}
H ~=~  \frac{p^2}{2 (\mu + (1 - x^2)^2)} + \frac{1}{4} x^4 - \frac{1}{2} x^2.
\end{equation}
Now in passing to the quantum theory we meet an ordering ambiguity, since $p$ and $x$ do not commute.   
We will adopt the ordering
\begin{eqnarray}\label{xHamiltonian}
H ~&=&~ - \frac{\hbar^2}{2} \,   \rho^{-\frac{1}{4}} \,  \frac{\partial}{\partial x}  \, \rho^{-\frac{1}{2}} \, \frac{\partial}{\partial x} \, \rho^{-\frac{1}{4}}  + \frac{1}{4} x^4 - \frac{1}{2} x^2 \\
\rho ~&\equiv&~  \mu + (1 - x^2)^2 \label{zebra}
\end{eqnarray}
as a simple prescription which leads to a Hermitian Hamiltonian.  As a formal matter, we can vary the numerical value of $\hbar$, to reflect the relationships between other dimensional parameters we might have included (but did not)  in Eqn.\,(\ref{tXtalL}) and its descendants.  Small values of $\hbar$ will emphasize the potential terms over the kinetic (gradient) terms, and thus de-emphasize the importance of the commutation relations, giving us the semiclassical limit.  Large values of $\hbar$, conversely, take us into the deep quantum regime.  

The same Hamiltonian (with $\mu = 0$) was obtained in \cite{brustein} by treating the Lagrangian of Eqn.\,(\ref{tXtalL}) canonically, as a constrained system.

It is entirely practical to solve for the eigenvalues and eigenfunctions of Eqn.\,(\ref{xHamiltonian}) numerically.  (Indeed, the {\it Mathematica\/} command ``NDEigensystem'' makes it easy.) An important qualitative result that emerges is that the spectrum remains stable as $\mu \rightarrow 0$.  Thus quantum mechanics itself regularizes the time crystal singularity, as anticipated.  We can also see this striking result emerging directly, without detailed calculation, by considering the behavior of the potentially singular contributions to Eqn.\,(\ref{xHamiltonian}) explicitly.   With $\mu \rightarrow 0$ the dangerous operator is the first term, in its behavior near $x = \pm 1$.   Considering for definiteness $x \approx 1$, we find for the most singular behavior
\begin{equation}
\frac{p^2}{2 (\mu + (1 - x^2)^2)} ~\rightarrow~ -\frac{1}{8} (\frac{\partial}{\partial x})^2 (\frac {1}{1-x})^2 ~ \sim ~ -\frac{1}{8} (\frac{\partial}{\partial x^2})_{x = 1}^2.
\end{equation}
Thus, using an appropriate variable, we can see that it is not singular after all. (Note that the potential is also quadratic around this point.)

For a more quantitative test, we can compare the low-lying eigenvalues of Eqn.\,(\ref{xHamiltonian}) with the semiclassical result Eqn.\,(\ref{semiClassicalEigenvalues}) which, we recall, was derived directly from the time crystal Lagrangian.   Figure 4 displays a representative result.

\begin{figure}[h!]
\begin{center}
\includegraphics[scale=0.6]{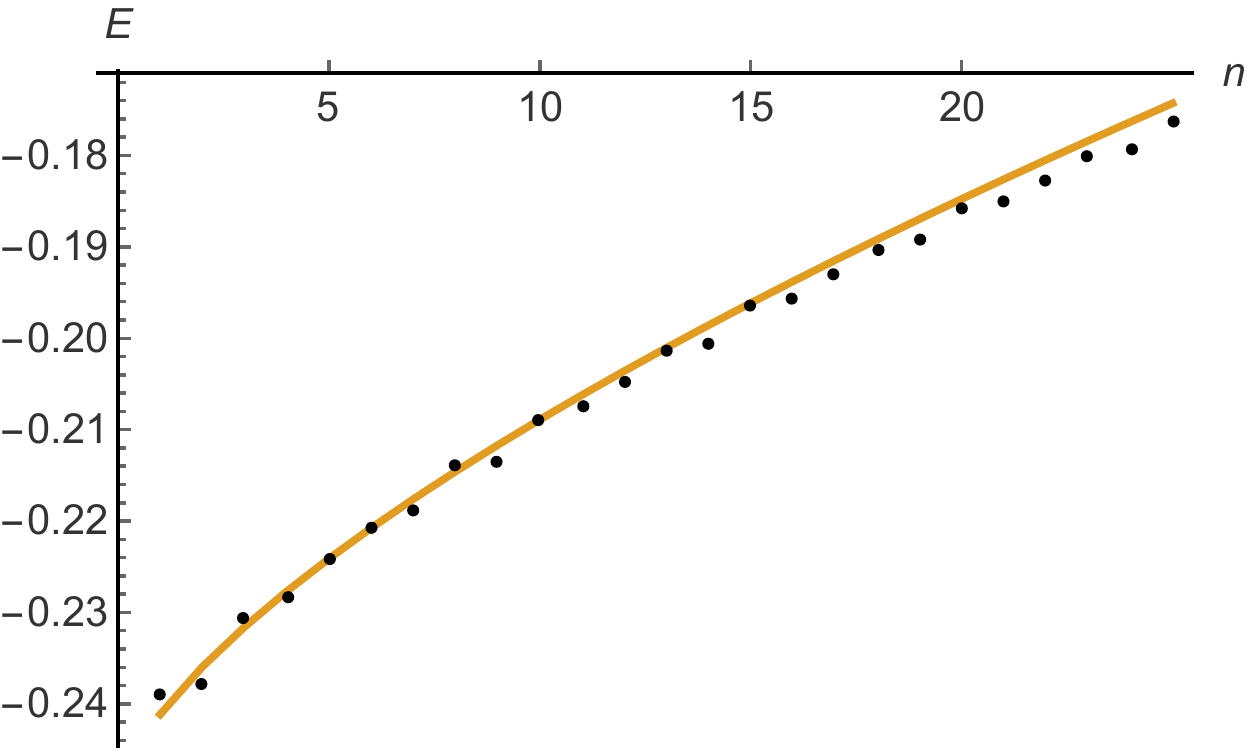}
\caption{Comparison of the parameter-free semiclassical predictions with numerically calculated values of the first twenty-five eigenvalues of Eqns.\,(\ref{xHamiltonian},\ref{zebra}) with $\hbar = .0005$, $\mu = 10^{-7}$.  The horizontal axis represents the level number $n$. The black dots are the numerical calculations; the gold curve is the semiclassical prediction of Eqn.(\ref{semiClassicalEigenvalues}).}
\end{center}
\end{figure}



\pagebreak

{\it Discussion\/} \ 
\begin{enumerate}
\item The new variable $x$ which is introduced in our regulation Eqn.\,(\ref{regulatedL}) of the time crystal Lagrangian enters through a unique term involving $\dot y$, which is linear in $\dot y$.   Thus it is a function of the momentum conjugate to $y$.  In particular, it does not introduce new degrees of freedom.  This should be contrasted with the (superficially) more straightforward approach of regulating singular behavior by adding higher derivatives \cite{muller1999,kohn2006}, i.e. 
\begin{equation}
\Delta L ~=~ \epsilon {\ddot y}^2 .
\end{equation} 
While that procedure manifestly overrides difficulties associated with vanishing coefficients in the highest-order terms in the equations of motion, it brings in other difficulties.  Besides implicitly introducing new degrees of freedom, it also introduces instabilities \cite{woodard2015}.  
\item We can introduce dissipation by adding a friction term $- \gamma \dot x$ on the right hand side of Eqn.\,(\ref{xEquation}).   With this addition, the system generically evolves toward the energy minimum.
\item As mentioned earlier, we can also encounter a form of Sisyphus dynamics in the purely spatial domain.  Indeed, consider a time-independent system governed by the Hamiltonian ($= -$ Lagrangian) density
\begin{equation}\label{staticL}
H ~=~ \frac{1}{12} u_x^4 + \frac{b}{6} u_x^3 + \frac{c}{2} u_x^2 + \frac{1}{2} u^2.
\end{equation}
We can regulate it using the same device as we used above for time crystal dynamics.   

Varying, we find the stress equation
\begin{eqnarray}
\partial_x T ~&=&~ 0 \\
T ~&\equiv&~ u_x^2 (\frac{1}{4} u_x^2 + \frac{b}{3} u_x + \frac{c}{2} ) - \frac{1}{2} u^2
\end{eqnarray}
For a given solution, let $\langle T \rangle$ be the constant value of $T$, and let us suppose that $u_{\rm max}^2$ is the maximum value of $u^2$.  (Since $u^2$ is energetically costly, the most interesting solutions are bounded in $u^2$.)  Then $f \equiv u_x^2 (\frac{1}{4} u_x^2 + \frac{b}{3} u_x + \frac{c}{2} ) $ satisfies 
\begin{equation}\label{fBand}
\langle T \rangle \leq f \leq \langle T \rangle + \frac{1}{2} u_{\rm max}^2
\end{equation} 

Now $f$, regarded as a function of $u_x$, defines the product of two parabolas. See Figure 5.  As is evident from that Figure, for some choices of the parameters $b, c, \langle T \rangle,  u_{\rm max}^2$,  the values of $u_x$ consistent with Eqn.\,(\ref{fBand}) will be confined to one or two small, positive intervals, leading unambiguously to Sisyphus dynamics.  For other values of $b, c$ the allowed intervals may support both positive and negative values of $u_x$.   See Figure 6. This opens up the possibility, commonly adopted in the calculus of variations and micro-materials literature \cite{muller1999,kohn2006}, to keep $u$ small by switching between positive and negative values of $u_x$, with appropriate joining prescriptions, e.g. using an $\epsilon u_{xx}^2$ regulator.    Our regulator suggests that Sisyphus dynamics, in effect allowing jumps in $u$ rather than $u_x$,  is a viable alternative here too.

\begin{figure}[h!]
\begin{center}
\includegraphics[scale=0.3]{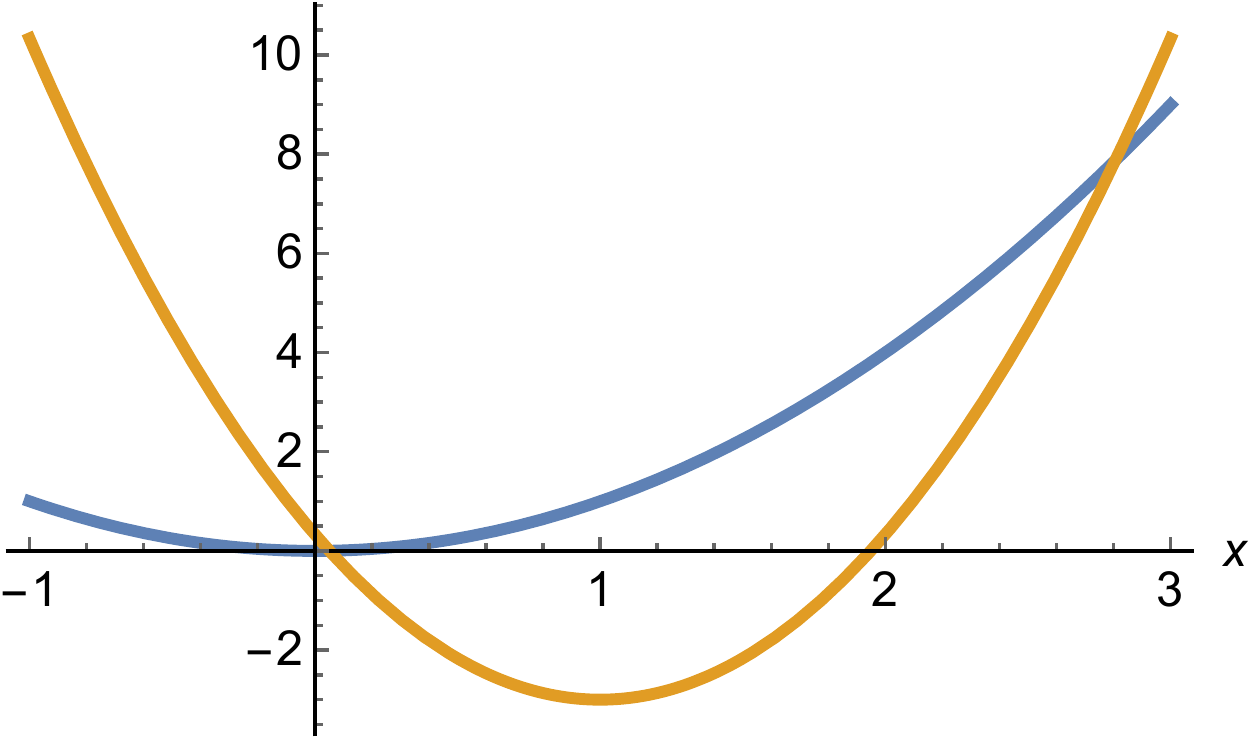}
\hskip.1in\includegraphics[scale=0.3]{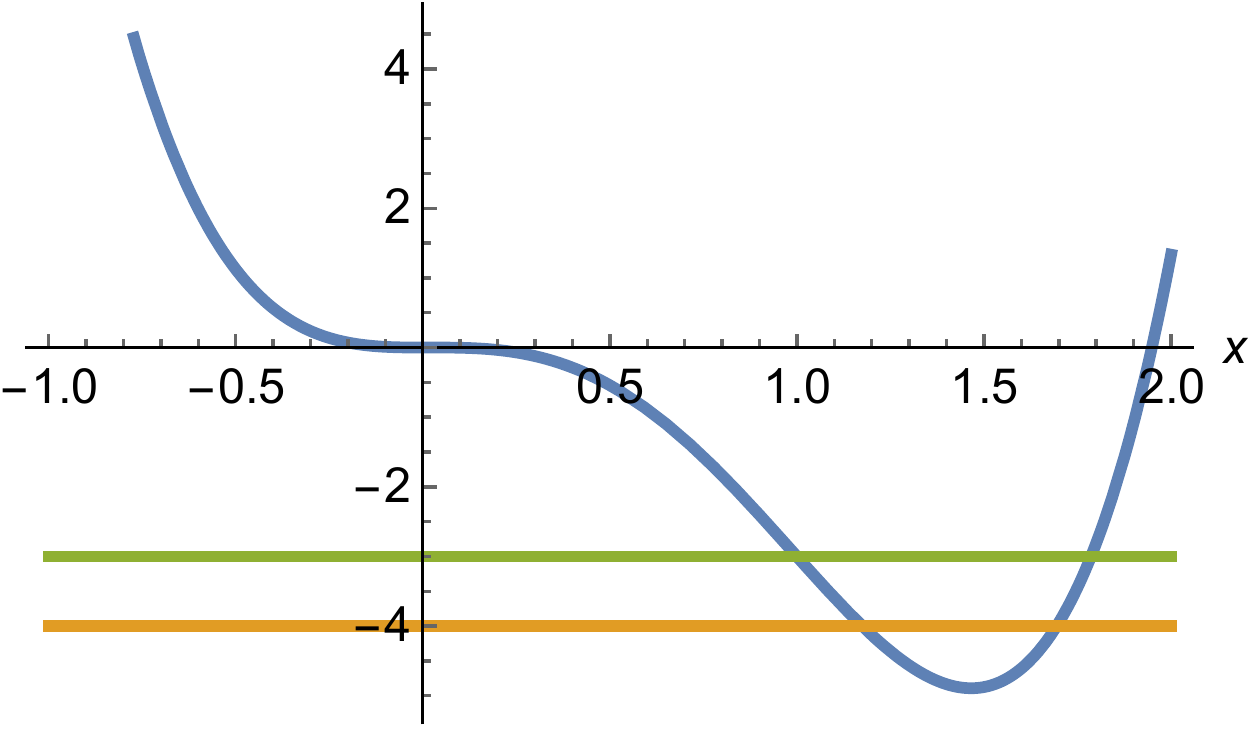}
\caption{On the left, two parabolas, one being simply a simple quadratic.   On the right, in blue, their product $f$ (defined above),  featuring a non-degenerate positive minimum.  When the value of $f$ lies between the green and orange horizontal lines, representing the upper and lower bounds in Eqn.\,(\ref{fBand}), only positive values of the abscissa $u_x$ appear between them.  As explained in the text, this leads to Sisyphus behavior. }
\label{positiveParabolas}
\end{center}
\end{figure}

\begin{figure}[h!]
\begin{center}
\includegraphics[scale=0.3]{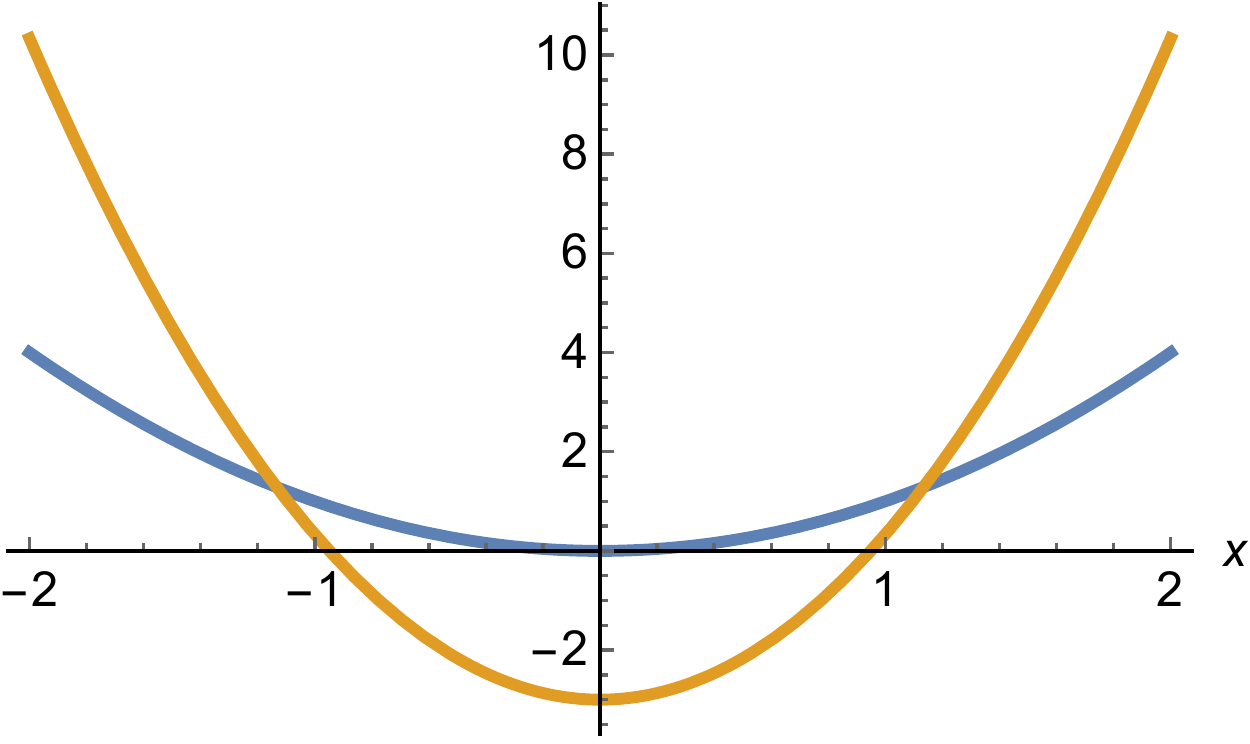}
\hskip.1in
\includegraphics[scale=0.3]{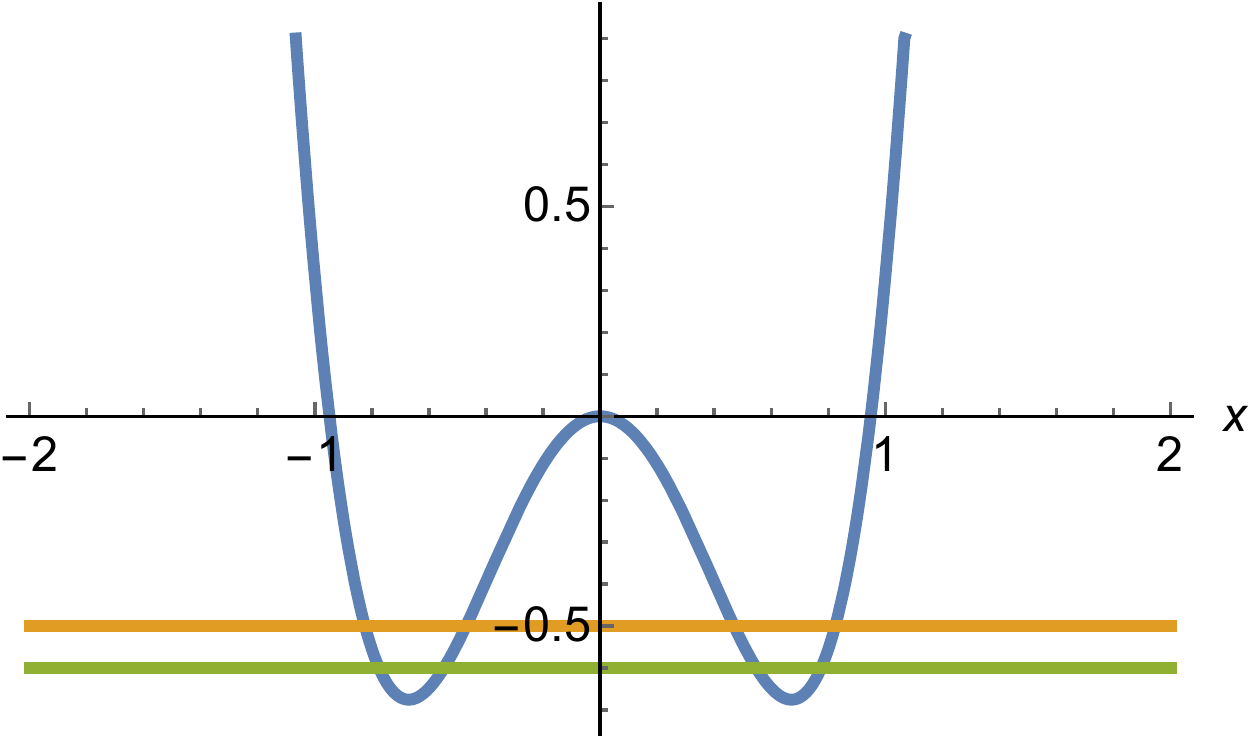}
\caption{Similar to Figure 5, but now both positive and negative values of the abscissa arise between the lines.}
\label{mixedParabolas}
\end{center}
\end{figure}

\item The Lagrangian $L = x({\dot y}^2 -a) - \frac{1}{2} x^2 - V(y)$ represents a kind of self-consistent effective mass for $y$, together with an effective potential.   Formally eliminating $x$ now leads to the reduced Lagrangian $ \frac{1}{2}( {\dot y}^2 -a)^2  - V(y)$, similar to what we had above.   This illustrates that the emergence of ``time crystal'' effective Lagrangians is more general than the specific model which we analyzed in detail above.

\item We can consider a variation on Eqn.\,(\ref{regulatedL}) using trigonometric functions. Taking $f = \frac{1}{3} \sin^3 x -\sin x, \, g = \frac{1}{4} \sin^4 x-\frac12 \sin^2 x$, and $\mu =0$, we recover Eqn.\,(\ref{tXtalL}).     One can insert an appropriate regulator, and a parallel analysis then applies.  Lagrangians of this sort describe periodic structures, and might also arise in the description of circuits including Josephson junctions.  Those possibilities merit further investigation.

\end{enumerate}

{\it Acknowledgements.\/} \  We thank Michael Weinstein and Robert V. Kohn for introducing us to the microstructure literature, and to Jordan Cotler for helpful discussions.  AS was partially supported by NSF grant PHY-1214341.  FW's work is supported by the U.S. Department of Energy under grant Contract  Number DE-SC0012567, by the European 
Research Council under grant 742104, and by the Swedish Research Council under Contract No. 335-2014-7424.

\end{document}